\documentclass[%
 aps,
 amsmath,amssymb,
 reprint,prmaterials%
]{revtex4-1}

\usepackage{graphicx}% Include figure files
\usepackage{dcolumn}% Align table columns on decimal point
\usepackage{bm}% bold math
\usepackage[utf8]{inputenc}
\usepackage[T1]{fontenc}
\usepackage{mathptmx}
\usepackage{xcolor}

\begin{document}

\preprint{AIP/123-QED}

\title{Strain game revisited for complex oxide thin-films:\\ Substrate-film thermal expansion mismatch in PbTiO$_3$}

\author{Ethan T. Ritz}
\affiliation{ 
Sibley School of Mechanical and Aerospace Engineering, Cornell University, Ithaca NY 14853}%
\author{Nicole A. Benedek}
\email{nbenedek@cornell.edu}
\affiliation{%
Department of Materials Science and Engineering, Cornell University, Ithaca NY 14853
}%

\date{\today}% It is always \today, today,
             %  but any date may be explicitly specified

\begin{abstract}
The sensitivity of materials properties, particularly those of perovskite oxides, to epitaxial strain has been exploited to great advantage to create materials with new or enhanced properties. Although it has certainly been recognized that mismatch in the thermal expansion coefficients of the bulk and substrate material will contribute to the misfit strain, the significance of this contribution for ferroelectric perovskite thin-films has not been systematically explored. We use first-principles density functional theory and the example of ferroelectric PbTiO$_3$ thin-films on various substrates to show that ignoring the thermal expansion of the substrate (that is, assuming that the in-plane lattice parameter of the film remains roughly constant as a function of temperature) results in ferroelectric transition temperatures and structural trends that are completely qualitatively different from calculations in which thermal expansion mismatch is properly taken into account. Our work suggests that the concept of a misfit strain defined as a single number is particularly ill-defined for PbTiO$_3$ and invites further study of the interplay between thermal expansion mismatch and structural and functional properties in other thin-film materials. 
\end{abstract}

\pacs{Valid PACS appear here}% PACS, the Physics and Astronomy
                             % Classification Scheme.
%Use showkeys class option if keyword
                              %display desired
\maketitle
Strain engineering has been an enormously successful technique for tuning the ground-state properties of materials. The particularly strong coupling between the polarization and strain in ferroelectric perovskites\cite{physicsferro,schlom2007strain,zhang09,schlom2014elastic,martin16} has been exploited to enhance the magnitude of the polarization in known ferroelectrics,\cite{choi04} induce ferroelectricity in nominally non-ferroelectric materials,\cite{haeni04} induce structural phase transitions in known ferroelectrics,\cite{zeches09,bousquet08} modify critical behavior\cite{kim19} and to create materials with multiple types of ferroic order.\cite{fennie06,lee10} Regardless of the material, or its application, perhaps the key parameter of interest in the strain game (as it is sometimes called\cite{schlom2014elastic}) is the misfit strain -- that is, the strain imparted to the growing film by the substrate due to a mismatch in the lattice constants of the substrate and film material. The misfit strain is typically quoted as a single number referenced to the lattice constants of the substrate and film material at some temperature, usually room temperature, or zero Kelvin in theoretical studies.

It has been noted in previous work\cite{janolin2007temperature,janolin09} that the internal stresses experienced by thin-films have multiple sources: lattice mismatch between the substrate and film material (as defined above), mismatch between the thermal expansion coefficients of the substrate and film material, and the existence of structural phase transitions. In addition to these intrinsic sources of stress, there are also contributions from extrinsic sources, such as point defects and dislocations. The structural parameters of the film, its thermal and other functional properties arise from a complex interplay between intrinsic and extrinsic stresses. Distinguishing between the effects of intrinsic and extrinsic stresses -- which is critical to both optimizing film growth techniques and to understanding the microscopic origins of thin-film properties -- is difficult and requires a combination of both careful experiments and theory.

The full range of theoretical techniques have been used to explore the properties of strained thin-films, from first- and second-principles techniques\cite{antons04,dieguez2005first,wojdel13,garcia16,sayalero17} and effective Hamiltonian methods parameterized for finite-temperature calculations,\cite{sepliarsky02,antons04,sepliarsky05,jiang14} to phenomenological models based on Ginzburg-Landau theory\cite{batra73,waser02,fong04,ahn04,fong06,li06,chandra07} and phase-field\cite{pertsev1998effect,li01,li02,li06} and finite-element models.\cite{hwang98} These techniques have yielded a wealth of information regarding, using the example of perovskites, the sequence of structural phases adopted as a function of temperature and ferroelectric domain morphologies. However, although the effects of misfit strain on transition temperatures, lattice parameters and domain morphologies are certainly well known from both theory and experiments, as far as we are aware, the question of how much thermal expansion mismatch between the film material and substrate affects thin-film properties has perhaps been less well-expolored.

We use density functional theory (DFT) and the quasiharmonic approximation (QHA) to reveal the critical role played by the mismatch in thermal expansion coefficients in giving rise to the structures of strained ferroelectric ($P4mm$) PbTiO$_3$ thin-films on SrTiO$_3$, DyScO$_3$ and (La$_{0.29(5)}$Sr$_{0.71(5)})_{\mathrm{A\,site}}$(Al$_{0.65(1)}$Ta$_{0.35(1)})_{\mathrm{B\,site}}$O$_3$ (LSAT) as a function of temperature. Ignoring the in-plane thermal expansion of the substrate results in structural and transition temperature trends that are completely qualitatively different from those in which thermal expansion mismatch is properly taken into account. Our work illustrates how the structural and thermal properties of thin-films can differ significantly from bulk even when the initial misfit strain is nominally zero (PbTiO$_3$ on SrTiO$_3$, for example) and provides a foundation for further development of recently proposed dynamical strain tuning strategies.\cite{tyunina2019perovskite,zhang2018continuously} Finally, the concept of the misfit strain as a single number seems particularly ill-defined for PbTiO$_3$, since two films with nominally the same lattice mismatch at a given temperature can evolve to have qualitatively different structures (and consequently, functional properties) depending on the thermal expansion of the substrates they are grown on.

\section{\label{sec:Methods}First-Principles Calculations}
All calculations were performed using density functional theory, as implemented in Quantum Espresso 5.3.0. \cite{giannozzi2009quantum} We used the Wu-Cohen exchange-correlation functional with Garrity-Bennett-Rabe-Vanderbilt ultrasoft pseudopotentials.\cite{garrity2014pseudopotentials} The following states were included in the valence for each element: 5d$^{10}$6s$^2$6p$^2$ for Pb, 3s$^2$3p$^6$4s$^2$3d$^1$ for Ti, and 2s$^2$2p$^4$ for O. Zero temperature unit cell lattice parameters and atomic positions of $P4mm$ PbTiO$_3$ were converged with respect to the plane wave cutoff energy and $\mathbf{k}$-point mesh density to within 0.001 \AA. Unless otherwise mentioned, structural parameters were found to be converged at a force cutoff threshold of $3.0 \times 10^{-5}$ Ry/bohr using an 8 $\times$ 8 $\times$ 8 Monkhorst-Pack (MP) mesh and an 80 Ry plane wave cutoff energy, compared with MP meshes up to 12$\times$12$\times$12 and plane wave cutoffs up to 90 Ry. Phonon dispersion calculations were performed using density functional perturbation theory (DFPT) on a 4 $\times$ 4 $\times$ 4 $\mathbf{q}$-point grid.

Finite temperature structural parameters were predicted using a quasiharmonic approximation (QHA) to the Helmholtz free energy -- in this study, we closely follow the framework outlined in Ref. \citenum{ritz2019thermal} regarding the application of the QHA. Our grid of strained systems consisted of a regular 7$\times$7 grid spanning -0.97\% to +2.02\% strain along the $a$-axis of the tetragonal phase, and -4.08\% to +1.98\% strain along the $c$-axis, augmented with 8 additional points with strains spanning -1.46\% to -0.47\% strain in $a$ and 0.97\% to +5.01\% strain in $c$. These points were chosen to ensure that for bulk PbTiO$_3$ and all strained films studied, the lattice parameters as a function of temperature would be bounded by these values. Note that the strain values above are defined with respect to the 0 K lattice parameters of the $P4mm$ phase found by minimizing the Helmholtz free energy ($a$ = 3.892 \AA, $c$ = 4.165 \AA), which includes the zero-point energy corrections from vibrational degrees of freedom. It is possible to define the misfit strain with respect to a difference reference, for example, to the $Pm\bar{3}m$ lattice parameter extrapolated to room temperature (with this reference, PbTiO$_3$ is under tensile strain on SrTiO$_3$, for example). We did investigate this approach as part of a study on the performance of phenomological models in predicting the temperature-dependent lattice parameters of PbTiO$_3$. However, we found that the results were quite sensitive to the details of how the extrapolation is performed. With respect to this study, since the cubic $Pm\bar{3}m$ phase exhibits unstable phonon modes at 0 K in a DFT calculation, its lattice parameters as a function of temperature cannot be predicted with the QHA, making it an impractical choice for our purposes. 

The QHA only explicitly includes the contributions of phonon-strain coupling to thermal expansion, while neglecting phonon-phonon contributions. However, extensive testing by us in a previous study\cite{ritz2018interplay} showed that the QHA qualitatively captures the changes in the lattice parameters of ferroelectric PbTiO$_3$, phonon frequency shifts, and changes in the elastic constants with temperature. See Refs. \onlinecite{ritz2018interplay} and \onlinecite{ritz2019thermal} for further details. For the calculations of epitaxially strained PbTiO$_3$, we clamp the in-plane ($a$ and $b$) lattice parameters of PbTiO$_3$ to the experimentally determined, temperature-dependent pseudocubic lattice parameters of the substrate, either SrTiO$_3$ (growth on the (100) surface), DyScO$_3$ (growth on (101) in the $Pbnm$ setting \cite{biegalski2006relaxor}) or LSAT (growth on the (001) surface).\cite{de1996high,biegalski2005thermal,chakoumakos1998thermal} That is, our misfit strain is defined as,
\begin{equation}
    \epsilon_a(T) = \frac{a_{\mathrm{substrate}}(T) - a_{\mathrm{bulk}}(T)}{a_{\mathrm{bulk}}(T)},
    \label{misfit}
\end{equation}
where $a_{\mathrm{substrate}}(T)$ is the lattice parameter of the substrate at some temperature $T$, and $a_{\mathrm{bulk}}(T)$ is the in-plane lattice parameter of bulk PbTiO$_3$ at the same temperature from our QHA calculations. Note that Equation \ref{misfit}, as written, is not completely general since it assumes that both the substrate and thin-film material can be described by a single pseudo-cubic lattice parameter. This is adequate not only for PbTiO$_3$ on the substrates considered here but also for most perovskites and common substrates. However, if there was significant anisotropy in either the substrate or thin-film material lattice parameters then a separate equation would be required to describe the misfit strain along each unique direction.

Since cubic $Pm\bar{3}m$ PbTiO$_3$ is unstable to both zone-center ferroelectric and zone-boundary distortions \cite{ghosez1999lattice}, we checked that the $P4mm$ phase is indeed the lowest in energy for the epitaxial strains considered in this work and is dynamically stable. We assume a monodomain film, coherent epitaxy, a uniform strain state throughout the film, and we ignore interfacial effects between the film and substrate.

\section{\label{sec:Results} Results and Discussion}
The misfit strain as defined in Equation \ref{misfit} contains the effects of both lattice mismatch and thermal expansion coefficient mismatch. As a first step, we would like to separately understand the effects of \emph{each} of these sources of stress on the lattice parameters and transition temperatures of strained ferroelectric PbTiO$_3$ thin-films. In other words, we would like to consider the misfit strain as a sum of two contributions,
\begin{equation}
    \epsilon_a(T) = \epsilon_a^{T = 300} + \epsilon_a^{\mathrm{thermal}}(T),
    \label{parts}
\end{equation}
where $\epsilon_a^{T = 300}$ is the intrinsic lattice mismatch defined with respect to some temperature (we use the usual convention of 300 K). Films also experience a temperature-dependent thermal stress due to the mismatch in in-plane thermal expansion coefficients between the bulk ($\alpha_a^{bulk}$) and substrate ($\alpha_a$). We denote the strain that arises from this source of stress as $\epsilon_a^{\mathrm{thermal}}$, where
\begin{equation}
\label{thermal}
 \epsilon_a^{\mathrm{thermal}}(T)=\int_{\tau=300}^T  \alpha_a(\tau)-\alpha_a^{bulk}(\tau)  \mathrm{d}\tau.
\end{equation}
Equation \ref{thermal} emphasizes that $\epsilon_a^{\mathrm{thermal}}$ changes with temperature depending on the thermal expansion coefficients of the bulk and substrate material.

\begin{figure}
    \centering
    \includegraphics[width=7cm]{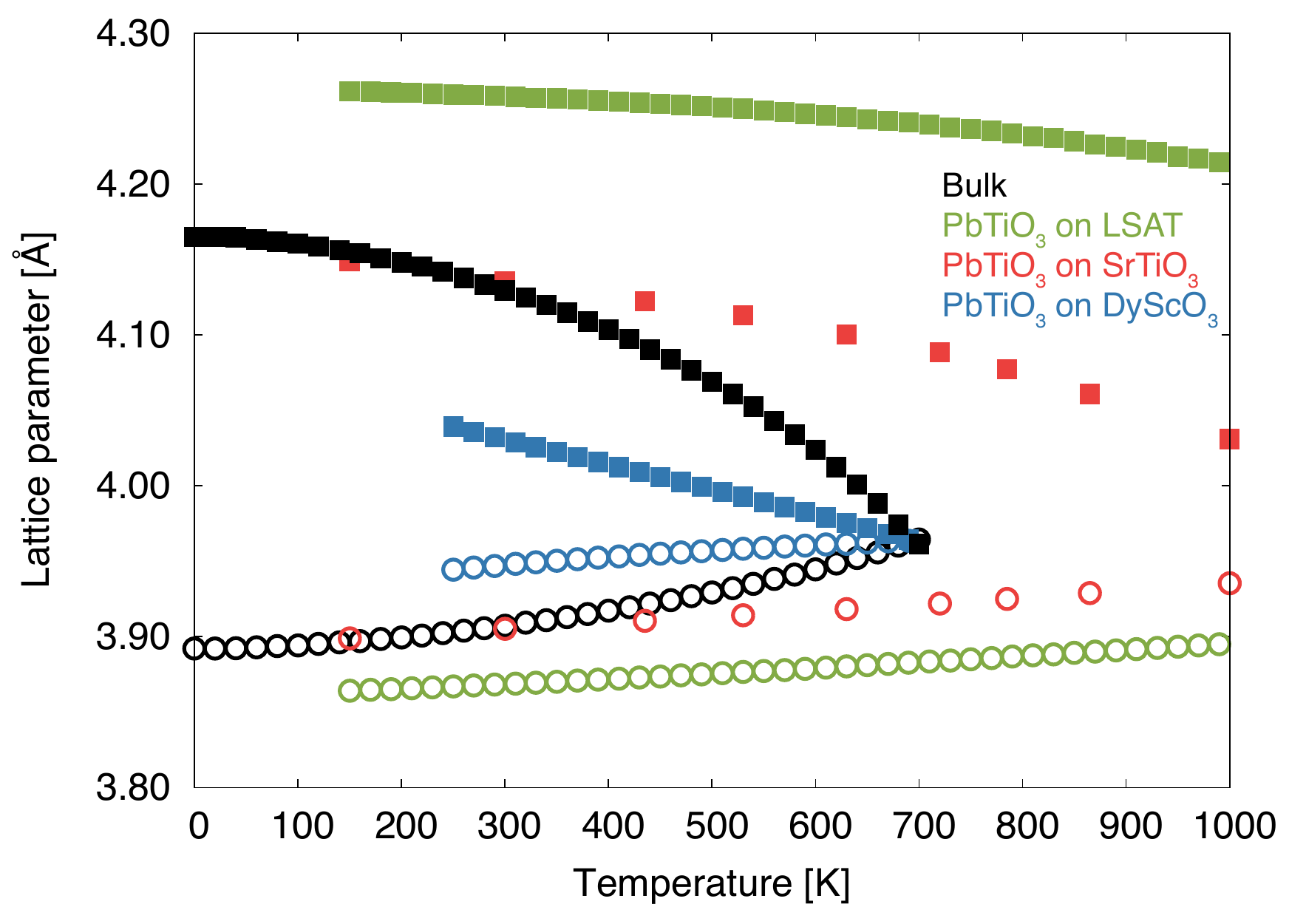}
    \caption{Variation in lattice parameters of PbTiO$_3$ thin-films with temperature and substrate compared to bulk PbTiO$_3$ from our first-principles calculations. Open circles represent data for the $a$-axis (experimental data for SrTiO$_3$, DyScO$_3$ and LSAT\cite{biegalski2005thermal,de1996high,chakoumakos1998thermal}), whereas closed squares represent data for the $c$-axis (predicted from our QHA calculations for each value of $a$).}
    \label{strain}
\end{figure}

In bulk, PbTiO$_3$ undergoes a phase transition from a cubic $Pm\bar{3}m$ phase to a ferroelectric $P4mm$ phase at $\sim$760 K. Figure \ref{strain} shows how the $c$-axis lattice parameters and ferroelectric phase transition temperatures of PbTiO$_3$ thin-films on various substrates change as a function of temperature. When the effects of both lattice mismatch and thermal expansion coefficient mismatch are included in our calculations (both terms in Equation 2), Figure \ref{strain} shows that $T_c$ increases compared to bulk for films grown on LSAT and SrTiO$_3$, whereas it decreases slightly for films grown on DyScO$_3$. In all cases, the $c$-axis shrinks with temperature, it just does so at a different rate compared to bulk, depending on the substrate. These trends align with conventional expectations, based on results from many decades of experimental and theoretical studies, regarding the relationship between misfit strain and $T_c$ for this system. The compressive strain imparted by LSAT makes PbTiO$_3$ more tetragonal, pushing it further away from the ferroelectric phase transition and increasing $T_c$. In contrast, the tensile strain imparted by DyScO$_3$ makes PbTiO$_3$ more cubic, pushing it closer to the ferroelectric phase transition and decreasing $T_c$. PbTiO$_3$ on SrTiO$_3$ is in between these two substrates and $T_c$ is accordingly between that of PbTiO$_3$ on DyScO$_3$ and PbTiO$_3$ on LSAT. This approximate relationship between strain state and $T_c$ is just a general rule of thumb but it has nonetheless been very useful.

\begin{figure}
    \centering
    \includegraphics[width=7cm]{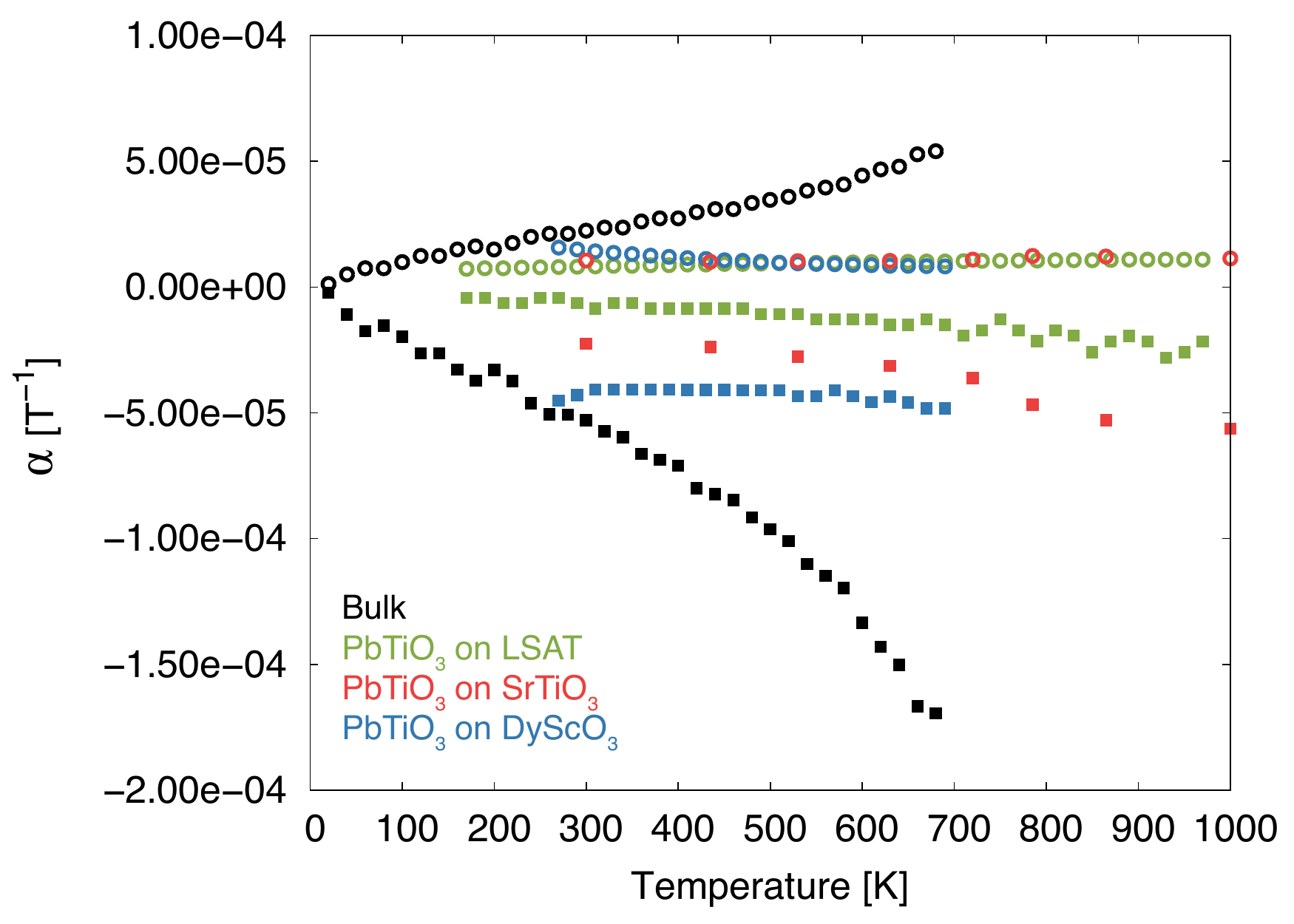}
    \caption{Variation in thermal expansion coefficients of PbTiO$_3$ thin-films with temperature and substrate compared to bulk PbTiO$_3$ from our first-principles calculations. Open circles represent data for thermal expansion along the $a$-axis ($\alpha_a$), whereas closed squares represent data for thermal expansion along the $c$-axis ($\alpha_c$). For the PbTiO$_3$ thin-films, the $\alpha_a$ values are those of the substrates and have been extracted from experimental data.\cite{biegalski2005thermal,de1996high,chakoumakos1998thermal} All $\alpha_c$ values were calculated for this work using the QHA.}
    \label{alphas}
\end{figure}

\begin{figure}
    \centering
    \includegraphics[width=6.5cm]{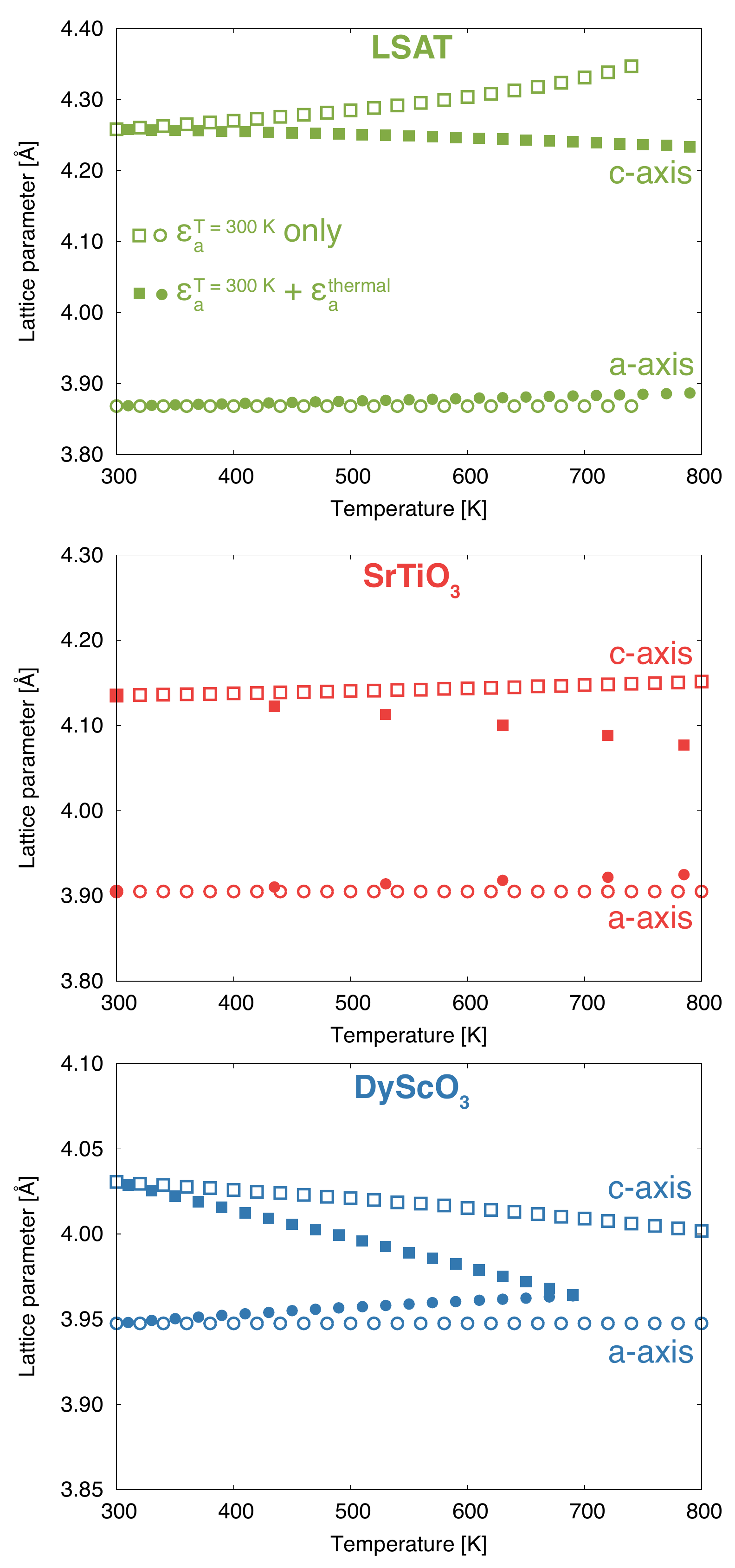}
    \caption{Variation in lattice parameters of PbTiO$_3$ thin-films with temperature when the lattice mismatch is held fixed at its 300 K value (open symbols, $\epsilon_a^{T = 300 K}$ term of Equation \ref{parts}) compared with lattice parameters calculated allowing for thermal expansion of the substrate (closed symbols, $\epsilon_a^{T = 300 K} + \epsilon_a^{\mathrm{thermal}}$). Squares denote $c$-axis lattice parameters from our QHA calculations, circles denote $a$-axis lattice parameters (experimental data\cite{biegalski2005thermal,de1996high,chakoumakos1998thermal}).}
    \label{substrates}
\end{figure}

Now, we would like to consider the role of the substrate thermal expansion coefficient in giving rise to the properties of PbTiO$_3$ thin-films. Figure \ref{alphas} actually shows that $\alpha_a$ for each substrate considered here is (coincidentally) nearly constant at about 1$\times$10$^{-5}$K$^{-1}$. In other words, the second term of Equation \ref{parts} ($\epsilon_a^{\mathrm{thermal}}$) is roughly the same for each substrate. What this means is that, in the case of PbTiO$_3$ thin-films on these particular substrates, it is the first term, the intrinsic lattice mismatch ($\epsilon_a^{T = 300 K}$), that is responsible for the difference in the structural and thermal properties of the thin-films grown on different substrates. However, we must be very careful not to interpret this result as meaning that the second term in Equation \ref{parts} does not matter -- it may be the same for the different substrates, but it is not zero. Figure \ref{substrates} shows how structural parameters and transition temperatures evolve if the in-plane lattice parameter of the substrate is held constant at its 300 K value compared with data for which the thermal expansion of the substrate is taken into account. Each panel contains a set of data for a single film-substrate system from Figure \ref{strain} (represented with filled symbols), alongside a fictitious system where the in-plane lattice parameter is fixed at the 300 K value of that substrate (represented with open symbols). For each panel, the first term in Equation \ref{parts} is the same for both the true film-substrate system (filled symbols) and the fictitious system with fixed in-plane lattice parameters (open symbols), but the second term in Equation \ref{parts} is different.  In the case of thin-films grown on LSAT and SrTiO$_3$, ignoring the effects of substrate thermal expansion causes the $c$-axis to grow with temperature, that is, the transition into the paraelectric phase is suppressed and the system remains ferroelectric. In the case of DyScO$_3$, $T_c$ remains finite but is pushed to much higher temperatures (far above bulk $T_c$) when only the intrinsic lattice mismatch is considered. What these results illustrate is that, in the case of PbTiO$_3$ thin-films on the substrates considered here, our intuition regarding the relationship between strain state and $T_c$ aligns with experimental observations \emph{only} because these substrates happen to have the `right' thermal expansion coefficients. If $\alpha_a$ was different, then we could expect \emph{qualitatively different} behavior, as we now discuss.

\begin{figure}
    \centering
    \includegraphics[width=9cm]{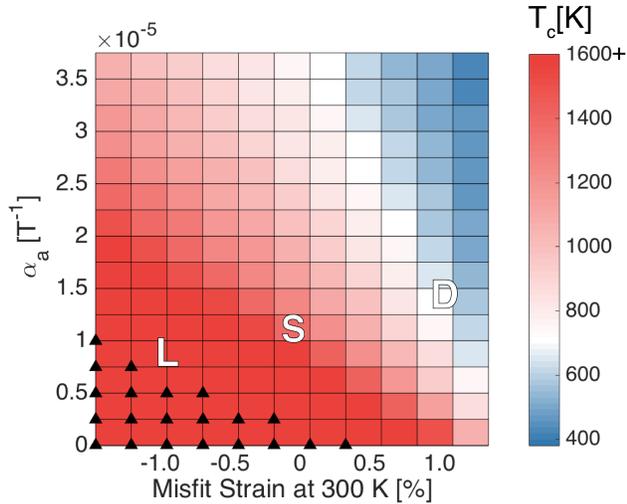}
    \caption{Variation in ferroelectric transition temperature and $c$-axis behavior of ferroelectric PbTiO$_3$ thin-films as a function of misfit strain and substrate coefficient of thermal expansion ($\alpha_a$) from our QHA calculations. The color chart for $T_c$ has white set to the bulk $T_c$ of 760 K. Blue (red) squares indicate a strain and $\alpha_a$ combination that produce films with a lower (higher) $T_c$ than bulk. Triangles indicate combinations of strain and $\alpha_a$ that produce films in which the $c$-axis continues to grow with temperature and $T_c$ is suppressed. All other combinations of strain and $\alpha_a$ produce films in which the $c$-axis shrinks with temperature and $T_c$ is finite. Note that the films with the highest transition temperatures may exceed 1600 K or even be completely suppressed. The letters `L', `S' and `D' denote the strain and $\alpha_a$ conditions corresponding to growth on LSAT, SrTiO$_3$ and DyScO$_3$.}
    \label{heatmap}
\end{figure}

Our results so far indicate that for the particular substrates considered here, $\alpha_a$ is roughly the same for each substrate and so $\epsilon_a^{\mathrm{thermal}}$ of Equation \ref{parts} is roughly the same. The properties of PbTiO$_3$ thin-films on SrTiO$_3$ (say) are different to those of PbTiO$_3$ on DyScO$_3$ because the initial lattice mismatch is different. The initial lattice mismatch does not just change the properties of thin-film materials, \emph{it changes how they change with temperature}. For a given initial lattice mismatch, however, what would the effect of changing $\alpha_a$ be? That is, if we had two different substrates that imparted the same strain at 300 K but different $\alpha_a$, how would the properties of the films on those two substrates differ? The answer to this question is revealed by Figure \ref{heatmap}, which shows how $T_c$ varies as a function of lattice mismatch at 300 K and substrate thermal expansion coefficient ($\alpha_a$, chosen to be constant with temperature) for a series of fictitious substrates. For a misfit strain of 1\% tensile at 300 K, for example, $T_c$ can either be higher than bulk or lower than bulk, depending on $\alpha_a$. For compressive strains and low or zero $\alpha_a$, the $c$-axis continues to grow with temperature such that the film remains ferroelectric (denoted by triangles). However, as $\alpha_a$ increases (moving vertically up the plot), the $c$-axis instead shrinks with temperature and although $T_c$ is higher than bulk, there is a transition into the paraelectric phase. What these results mean is that there is no reason to expect that thin-films of the same material grown on two different substrates that impart the same lattice mismatch at 300 K, but have different thermal expansion coefficients, should have the same structures or functional properties. Indeed, the effect of the thermal expansion coefficient of the substrate on thin-film properties is striking and cannot be over-stated: taken together, Figures \ref{substrates} and \ref{heatmap} essentially show that the thermal expansion properties of the substrate have \emph{at least} as large an effect on thin-film structural and functional properties as the intrinsic lattice mismatch. 

\begin{figure}
    \centering
    \includegraphics[width=6cm]{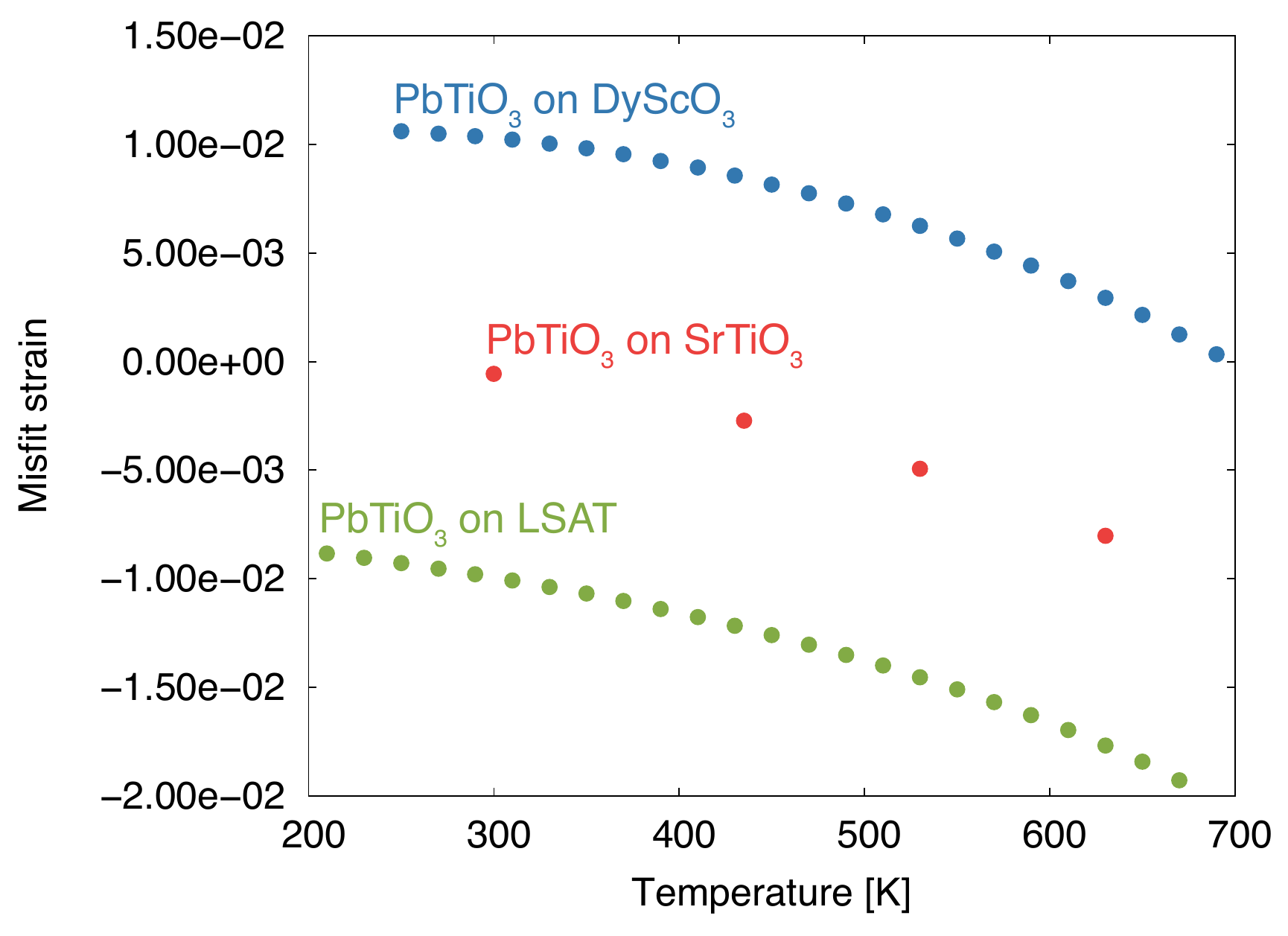}
    \caption{Variation in in-plane misfit strains as a function of temperature for PbTiO$_3$ thin-films on DyScO$_3$, SrTiO$_3$ and LSAT from our QHA calculations.}
    \label{misfits}
\end{figure}

Why does the thermal expansion coefficient of the substrate have such a significant effect on the properties of ferroelectric PbTiO$_3$ thin-films? How much does it change the misfit strain as a function of temperature? Figure \ref{misfits} shows how the misfit strain varies with temperature for PbTiO$_3$ thin-films on LSAT, SrTiO$_3$ and DyScO$_3$. In the case of SrTiO$_3$, the lattice match is almost perfect at 300 K, but the misfit strain increases to almost 1\% above 600 K. In the case of LSAT and DyScO$_3$, the misfit strain fully doubles from around 1\% at 300 K, to around 2\% above 600 K (compressive for LSAT, tensile for DyScO$_3$). These enormous changes in misfit strain can be expected to dramatically change the properties of PbTiO$_3$ thin-films as a function of temperature, an especially important consideration in practical applications that involve temperature cycling. The effect may be particularly pronounced for PbTiO$_3$ thin-films because the thermal expansion coefficient of the $a$-axis for PbTiO$_3$ is so large and increases with temperature, as Figure \ref{alphas} shows. In contrast, $\alpha_a$ for each substrate is nearly constant at about 1$\times$10$^{-5}$ K$^{-1}$, as we have already discussed. Hence, as the temperature increases, the substrates considered here force the PbTiO$_3$ film to expand much slower along its in-plane axis than it would normally at a given temperature. Figures \ref{alphas} and \ref{misfits} also explain why the temperature-dependent structural properties of PbTiO$_3$ thin-films on SrTiO$_3$ differ so much from bulk, despite the near-perfect lattice match at 300 K. Although the lattice constants of SrTiO$_3$ and PbTiO$_3$ are very similar at 300 K, their in-plane ($\alpha_a$) thermal expansion coefficients are not, and the difference increases with temperature.

\begin{figure}
    \centering
    \includegraphics[width=6cm]{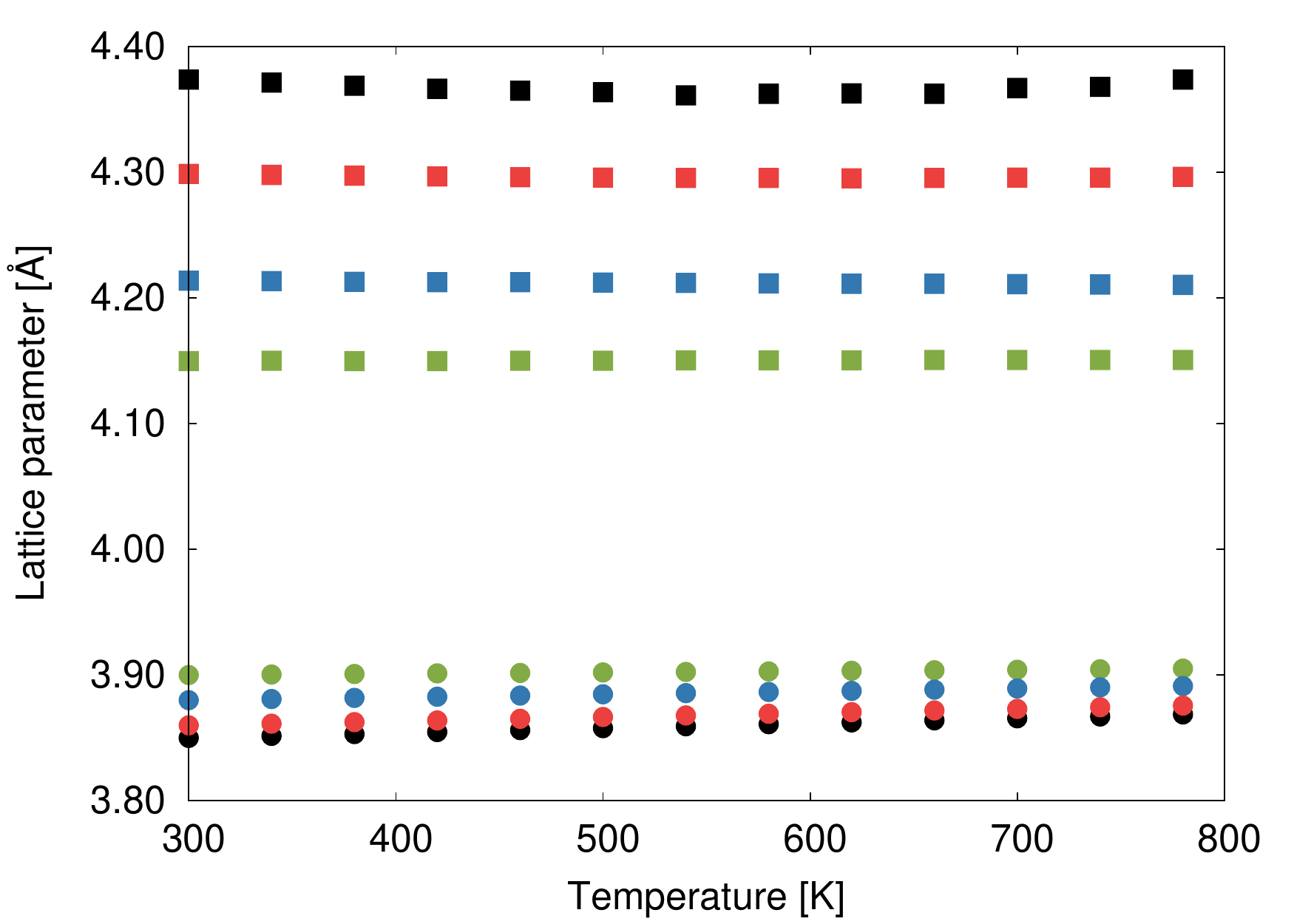}
    \caption{Judicious combinations of initial lattice mismatch (defined at 300 K) and substrate coefficient of thermal expansion ($\alpha_a$) can produce PbTiO$_3$ thin-films with zero thermal expansion along the $c$-axis with temperature. Circles represent data for the $a$-axis whereas squares represent data for the $c$-axis. Black: -1.46\% misfit strain (300 K), $\alpha_a$ = 1.01$\times$10$^{-5}$ K$^{-1}$. Red: -1.21\% misfit strain, $\alpha_a$ = 0.85$\times$10$^{-5}$ K$^{-1}$. Blue: -0.70\% misfit strain, $\alpha_a$ = 0.60$\times$10$^{-5}$ K$^{-1}$. Green: -0.18\% misfit strain, $\alpha_a$ = 0.27$\times$10$^{-5}$ K$^{-1}$.}
    \label{invar}
\end{figure}

Finally, we consider some of the broader implications of our results. Although all the substrates we considered have roughly the same, fairly low, thermal expansion coefficients, we can imagine how the structural and functional properties of a thin-film my be tuned via judicious choice of intrinsic lattice and thermal expansion coefficient mismatch. Figure \ref{invar} shows the change in the $c$-axis lattice parameter as a function of temperature for growth of PbTiO$_3$ thin-films on four different (fictitious) substrates with different intrinsic lattice mismatch and different $\alpha_a$. For the particular combinations shown in Figure \ref{invar}, thermal expansion of the $c$-axis is suppressed, leading to thin-films with Invar-like behavior. It should also be possible to engineer thin-films with greatly enhanced positive or negative thermal expansion by appropriate choice of substrate. Recent work has also demonstrated the potential for exploiting thermal expansion mismatch to achieve continuous strain tuning. For example, Zhang and co-workers showed that the amount of tensile strain imparted to thin-films of SrTiO$_3$ grown on a Si substrate could be continuously tuned by varying the growth temperature.\cite{zhang2018continuously,hellberg06} In this study, the SrTiO$_3$ thin films are allowed to relax to their unstrained bulk lattice parameters at the growth temperature, owing to the formation of dislocations and an amorphous SiO$_x$ layer at the interface between the film and substrate. Then, since the thermal expansion coefficients of SrTiO$_3$ and Si are quite different, and the SrTiO$_3$ film is clamped to the Si substrate through the amorphous oxide layer,  changes in the lattice parameter of Si with temperature upon cooling the system to room temperature results in a residual tensile strain in the SrTiO$_3$; the magnitude of this strain depends on the growth temperature. It would be particularly interesting to employ this strategy on a system that undergoes a phase transition in range of the growth temperature, to compare films relaxed above and below the phase transition. Our work helps build a theoretical basis by which to predict how this process, and other strategies for dynamical tuning of structural properties using thermal strain\cite{tyunina2019perovskite}, could affect the structural properties of other combinations of films and substrates.

\section{\label{sec:Conclusion} Conclusions}
The results of our first-principles study of PbTiO$_3$ thin-films show that \emph{both} the intrinsic lattice mismatch and the thermal expansion coefficient mismatch play critical roles in determining how the structures and functional properties of PbTiO$_3$ thin-films evolve with temperature. We showed that when the in-plane lattice parameters of the substrate is held constant and the only difference between thin-films on different substrates is the intrinsic lattice mismatch at 300 K, the evolution of the $c$-axis with temperature is qualitatively different and can even change sign compared to results when the thermal expansion is taken into account. The intrinsic lattice mismatch does more than just specify the `initial conditions' of the lattice parameters however, it also has important implications for how the lattice parameters of the film will change with temperature. Correctly capturing these responses is critical for understanding not only how the structural properties of the film evolve with temperature but also all the properties linked to the $c/a$ ratio, such as the ferroelectric polarization. We further showed how transition temperatures and structural properties can qualitatively differ based on the thermal expansion coefficient of the substrate, even when the intrinsic lattice mismatch at 300 K is constant. While we expect that the thermal expansion mismatch plays an important role in giving rise to the structural properties of thin-films of materials other than PbTiO$_3$, the magnitude of the effect will of course be different. This is because the materials properties most likely to control the microscopic mechanisms underlying the behavior discussed here -- elasticity and vibrational effects -- are material dependent. Understanding the microscopic mechanisms that drive these behaviors is an exciting avenue for future work, and we plan to explore these details in a forthcoming publication.

\begin{acknowledgments}
This work was supported by the National Science Foundation. E. T. R. and N. A. B. were supported by DMR-1550347. Computational resources were provided by the Cornell Center for Advanced Computing and the Extreme Science and Engineering Discovery Environment (XSEDE) through allocation DMR-160052. The authors thank Jorge \'{I}\~{n}iguez, Craig Fennie, and Darrell Schlom for helpful discussions.
\end{acknowledgments}

\bibliography{citations}

\end{document}